%% file: sollerman_main.tex
\documentclass[multphys,vecphys]{svmult}

\usepackage{makeidx}     % allows index generation
\usepackage{graphicx}    % standard LaTeX graphics tool
\usepackage{multicol}    % used for the two-column index

\makeindex             % used for the subject index
\begin{document}

\title*{The late UVOIR light curve of SN 2000cx}
% Use \titlerunning{Short Title} for an abbreviated version of
% your contribution title if the original one is too long
\author{Jesper Sollerman\inst{1}\and
Cecilia Kozma\inst{1}\and
Jan Lindahl\inst{1,2}
}
% Use \authorrunning{Short Title} for an abbreviated version of
% your contribution title if the original one is too long
\institute{Stockholm Observatory, AlbaNova, 106 91 Stockholm, Sweden
\texttt{jesper@astro.su.se}
\and Onsala Space Observatory, 439 92 Onsala, Sweden}
%
% Use the package "url.sty" to avoid
% problems with special characters
% used in your e-mail or web address
%
\maketitle
We present preliminary data and modeling of the late time light curve of the Type Ia supernova SN 2000cx. Optical and near-infrared data obtained with the VLT at 360 to 480 days past maximum light show the increasing importance of the near-infrared regime. Detailed multi-band modeling based on W7 also show this effect. Conclusions on positron escape in this phase may therefore require more detailed observations and modeling than hitherto appreciated.

\section{Introduction}
\label{intro}
% Always give a unique label
% and use \ref{<label>} for cross-references
% and \cite{<label>} for bibliographic references
% use \sectionmark{}
% to alter or adjust the section heading in the running head

Type Ia supernovae (SNe Ia) are believed to be the destructive thermonuclear
explosions of white dwarfs.
% (e.g., Hoyle \& Fowler 1960; Leibundgut 2000). 
Their light curves are during the first couple of years powered by the
radioactive decay of freshly synthesized $^{56}$Ni, releasing
$\gamma$-rays and positrons in the ejecta. 
At phases later than about 200 days, virtually all
gamma-rays escape freely from the ejecta, and the luminosity is
then provided by the kinetic energy deposited by the positrons. 

Whether or not the positrons are able to slip out of
the ejecta depends on the strength and geometry of the magnetic field
(e.g., \cite{ps98,m99,m01}).
%Ruiz-Lapuente \& Spruit 1998; Milne et al. 1999,2001). 
While a weak and radially combed magnetic field might allow positron escape,
thus providing a steep light curve, a strong, tangled magnetic field
would efficiently trap all the positrons, and drive the light curve to
the radioactive decay rate. 

As a first step to investigate whether or not observations of
the positron phase can establish conclusions about positron escape, 
we have conducted a photometric study at late phases of the
SN Ia 2000cx, and modeled its light curve in detail.
Here we present some preliminary results from that study.
The final analysis will be reported elsewhere.

\section{Observations of SN 2000cx}

SN 2000cx was discovered on 17.5 July 2000 \cite{ym00} far from
%and was located $23\farcs0$ west and
%$109\farcs3$ south of 
the nucleus of the S0 galaxy NGC 524, and became the 
brightest supernova
observed that year. 
This made it a very
%Furthermore, the favorable position far from the
%galaxy center made it a 
good target for late time photometry. The early
evolution has been extremely well covered \cite{li01,cand03}.

We have observed the field of SN 2000cx in the optical, (U)BVRI, 
regime during four epochs between 360 and 480 days past maximum light.
These observations were obtained with the FORS instruments at the ESO VLT.
The data were reduced in a standard way using {\tt IRAF}. 

Near-infrared observations were obtained using the ISAAC instrument at the VLT.
Data were obtained in the J and H (and K) 
bands at three epochs close to the optical 
observations. The data were reduced within {\tt Eclipse} and {\tt IRAF}. 

Magnitudes where measured using 
aperture photometry and zeropoints in the standard system were obtained by 
observations of standard stars.

A detailed description of the observations and the data reductions 
will be given elsewhere. Very late observations using the HST will be 
included in a future study.

\section{Modeling}

We have performed detailed modeling of the emission from SNe Ia
in the nebular phase, 250 - 1000 days after explosion, in order to interpret 
our observations of SN 2000cx. 
The code is an updated version of the code described by Kozma \& Fransson \cite{kf98a}.
% and will be described in detail elsewhere.

The decay of $^{56}$Co dominates the energy input at the epochs we are modeling. 
The gamma-rays emitted in the decays 
give rise to fast electrons which 
deposit their energy by heating, ionizing, or exciting the ejecta. 
In our calculations we assume full and immediate positron deposition 
within the regions containing the newly synthesized iron.

As input to our calculations we use the density structure, 
abundances and velocity 
structure from model W7 \cite{n84,t86}.
UV-scattering is not included and as it may be important for the ionization structure
%To probe the effects of photoionization due to UV-scattering 
we have 
made model calculations with and without including photoionization.
%In the model including photoionization we find that the ejecta is 
%photoionized mainly by recombination emission.

\section{Results}

In Fig.~1 the light curves for the $B$- to $H$-bands
are shown for both observations and models.
We find a general good agreement between
observations and models, with a steeper slope in the $BVR$-
(V band declines 1.4 mag. per 100 days) 
and virtually constant light curves in the $JH$-bands.

The curves for the model without 
photoionization (dotted) drops quickly after 400-600 days. 
This is due to a lower temperature in the ejecta for this model. 
%without photoionization. The higher temperatures in the photoionization
%model results in a less steep decrease in the light curves 
For the $BVRI$-bands the
model including photoionization (dashed-dotted) 
appears to give a better fit to the observations.
However, since the differences in the light curve are mainly
a temperature effect, other processes might give
the same result. For example, clumping of the ejecta would 
allow low density regions to be hotter, keeping
the light curves from dropping. 

\begin{figure}
\centering
% Use the relevant command for your figure-insertion program
% to insert the figure file.
% For example, with the option graphics use
%\includegraphics[height=12cm]{phot.00cx.012_013_bvrijh.epsi}
\includegraphics[height=12cm]{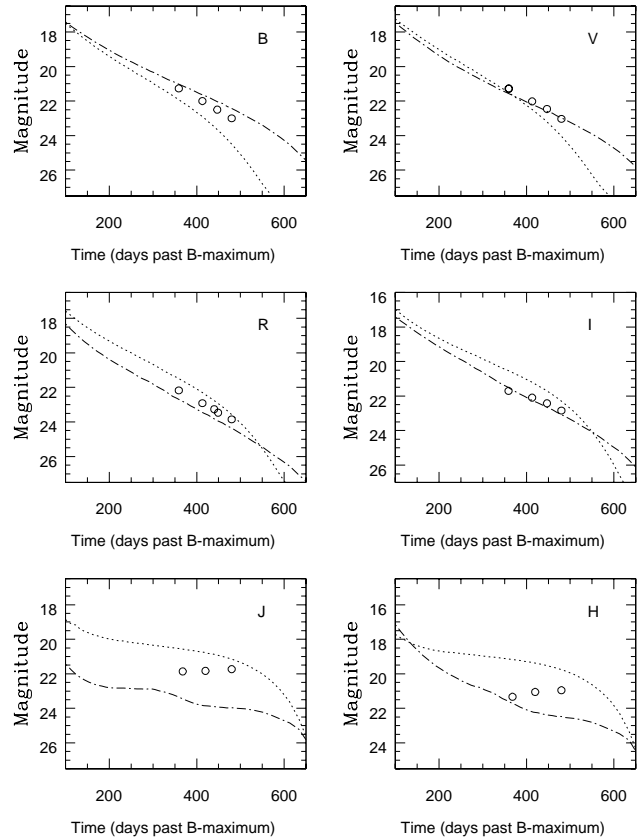}
%
% If not, use
%\picplace{5cm}{2cm} % Give the correct figure height and width in cm
%
\caption{Late light curves of SN 2000cx. 
The 
%observational 
errorbars fall within the circles.
}
\label{fig:1}       % Give a unique label
\end{figure}

\subsection{Bolometric luminosity}

The wide coverage of broad band magnitudes allows an attempt to construct a uvoir 'bolometric' lightcurve.
We estimated the bolometric lightcurve by simply integrating the flux from (U)B to H(K) at all epochs. The result is shown in Fig.~2.

\begin{figure}
\centering
\includegraphics[height=10cm]{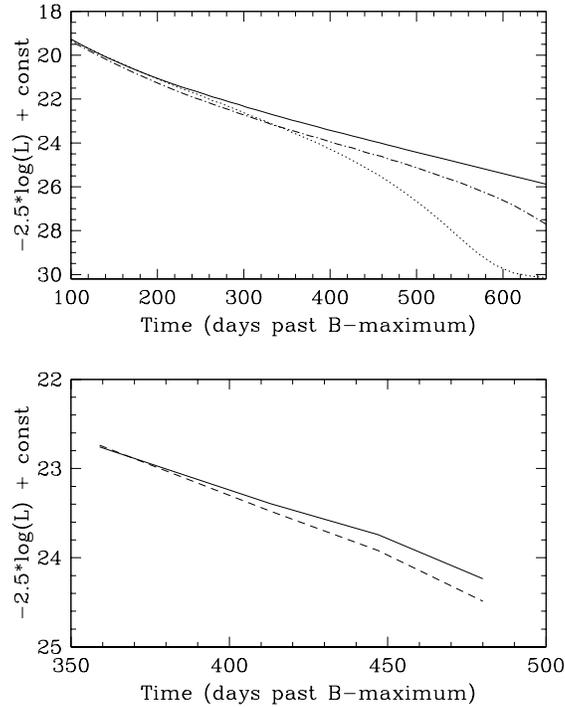}
\caption{Upper panel: The full curve shows the modeled bolometric luminosity
and the others show %the luminosity in 
the V-band for the two models (see Fig. 1). The curves are matched to the same value at 100 days. 
Lower panel: The full curve shows the observed integrated luminosity in the (U)BVRIJH(K) bands, while the dashed curve is the observed luminosity in the V-band. For both models and observations the slope of the 'bolometric' light curve diverges from the V-band light curve at later epochs.}
\label{fig:2}       % Give a unique label
\end{figure}

\section{Discussion}

As the trapping of the $\gamma$-rays decreases, the kinetic energy of the positrons will start to dominate the energy input to the ejecta.
Our model assumes full trapping of the positrons.
% which is likely to be the case for the ejecta of SNe
%Ia with a tangled magnetic field. 
In this case the bolometric light curve should
flatten out and approach the decay rate of $^{56}$Co in the positron
dominated phase.  However, late observations of SNe Ia in different
pass bands indicate that the light curve continues to fall more
rapidly also at epochs later than 250 days. 
This has been interpreted as due to positron escape \cite{capp97,m99,m01,ps98}.

However, since the observations of late light curves are sparse, in particular in the near-IR,  it is generally not possible to construct true bolometric light curves. 
For example, Cappellaro et al. \cite{capp97} had to assume that
the late bolometric light curve follows
the V-band. This assumption is not necessarily valid. As the
input heating decreases and the ejecta expands the temperatures will decrease, 
and color evolution could mimic the effect of positron escape.

In Fig. 2 we compare the bolometric light curve to the V-band 
light curve for our two model calculations. 
% and our observations. 
Even with full positron trapping we find an increasing 
deviation between the bolometric and V-band light curve
with time. Especially for the model without photoionization the
drop in the V-band is rapid around 400-500 days. 

The emission in the various bands behave quite differently, due to the evolution of temperature and ionization structure within the ejecta. This makes it hazardly to assume that any particular band reflects the true bolometric luminosity.
Also in the observations do we find that the slope of the bolometric and V-band light curves differs (Fig. 2 lower panel). A comparison to a true bolometric light curve would increase this effect.

We therefore find it difficult to draw any conclusions about 
the degree of positron escape in SNe Ia without having a detailed and consistent knowledge of the temperature and ionization evolution of the ejecta.
Time dependent bolometric corrections can instead be a likely explanation for these observations.\\

\noindent
{\em Acknowledgements.}

We are grateful to K. Nomoto for providing the W7 model,
and to S. Nahar for iron recombination rates.
We thank B. Leibundgut, C. Fransson, P. Lundqvist, N. Suntzeff and the SINS team, as well as P. Milne for discussions on this project.
The observations are based on observations collected at the European Southern Observatory (67.D-0134 and 68.D-0114).
JS trip to Valencia was paid by Wallenbergsstiftelsens jubileumsfond.

\input{sollerman_ref}

%%%%%%%%%%%%%%%%%%%%%%%%%%%%%%%%%%%%%%%%%%%%%%%%%%%%%%%%%%%%%%%%%%%%%%  }

%%%%%%%%%%%%%%%%%%%%%%%%%%%%%%%%%%%%%%%%%%%%%%%%%%%%%%%%%%%%%%%%%%%%%%

\printindex
\end{document}

%% file: sollerman_ref.tex
%%%%%%%%%%%%%%%%%%%%%%%% referenc.tex %%%%%%%%%%%%%%%%%%%%%%%%%%%%%%
% sample references
% "physics"
%
% Use this file as a template for your own input.
%
%%%%%%%%%%%%%%%%%%%%%%%% Springer-Verlag %%%%%%%%%%%%%%%%%%%%%%%%%%

%
% BibTeX users please use
% \bibliographystyle{}
% \bibliography{}
%
% Non-BibTeX users please use